\documentclass[fleqn,10pt]{wlscirep}
\usepackage[utf8]{inputenc}
\usepackage[T1]{fontenc}
\usepackage{lineno}
\usepackage{subcaption}
\usepackage{float}

\newcommand{\q}[1]{``#1''}

\title{ApisTox: a new benchmark dataset for the classification of small molecules toxicity on honey bees}

\author[1,$\dag$,*]{Jakub Adamczyk}
\author[2,$\dag$]{Jakub Poziemski}
\author[2]{Pawel Siedlecki}
\affil[1]{AGH University of Krakow, Department of Computer Science, Cracow, Poland}
\affil[2]{Institute of Biochemistry and Biophysics, Polish Academy of Sciences, Warsaw, Poland}

\affil[*]{corresponding author: Jakub Adamczyk (jadamczy@agh.edu.pl)}

\affil[$\dag$]{these authors contributed equally to this work}

\begin{abstract}
The global decline in bee populations poses significant risks to agriculture, biodiversity, and environmental stability. To bridge the gap in existing data, we introduce ApisTox, a comprehensive dataset focusing on the toxicity of pesticides to honey bees (Apis mellifera). This dataset combines and leverages data from existing sources such as ECOTOX and PPDB, providing an extensive, consistent, and curated collection that surpasses the previous datasets. ApisTox incorporates a wide array of data, including toxicity levels for chemicals, details such as time of their publication in literature, and identifiers linking them to external chemical databases. This dataset may serve as an important tool for environmental and agricultural research, but also can support the development of policies and practices aimed at minimizing harm to bee populations. Finally, ApisTox offers a unique resource for benchmarking molecular property prediction methods on agrochemical compounds, facilitating advancements in both environmental science and chemoinformatics. This makes it a valuable tool for both academic research and practical applications in bee conservation.
\end{abstract}
\begin{document}

\flushbottom
\maketitle

\thispagestyle{empty}

\section*{Background \& Summary}

Bees are essential to the pollination of plants, and protecting their populations is of crucial importance for environmental preservation and food security \cite{bees_essential_1,bees_essential_2}. Global declines of their population are a serious threat to agricultural production, environmental stability, and overall biodiversity \cite{bees_decline}. It is a complex issue influenced by many factors, including exposure to pesticides, other human introduced variables, and climate change \cite{bees_decline_reasons}. However, data on the effects of various crop management systems, as captured e.g. in CropCSM \cite{CropCSM} datasets, are often disconnected from bee toxicity information, hindering more holistic analyses. To evaluate the effects of different stressors, such as pesticides, infections, and environmental changes on bee health, thorough and trustworthy data is needed \cite{bee_decline_data_1,bee_decline_data_2}. This necessity motivated the creation of ApisTox, the largest single resource on honey bee (\textit{Apis mellifera}) toxicity data publicly available.

None of the  currently available databases cover the full range of pesticides and other toxic substances that bees are exposed to in real-world scenarios. In fact, bee toxicity data is scattered around many sources, encompassing both specialized ones (e.g. BeeTox \cite{BeeTox}), and general-purpose pesticide or ecotoxicology databases, such as PPDB \cite{PPDB}, BPDB \cite{PPDB}, and ECOTOX \cite{ECOTOX}. Existing datasets on bee toxicity offer valuable insights, but are hindered by many inconsistency problems. For example, toxicity data points exist in different formats, units, and organization across reference databases. Databases also frequently list substances not relevant to honey bees, requiring data filtering and curation. Apart from the data quality, problems occur also with records duplication or improper chemical information, which can even be unreadable for chemoinformatics programs and pipelines. Lack of additional data relevant for QSAR, such as formulation use or pesticide introduction date, is further limiting the scope, consistency, and comprehensiveness of currently available bee toxicity data.

To address some of these gaps, we have developed a comprehensive dataset that combines carefully filtered and manually curated data from the above-mentioned resources and augment it with additional relevant information. By consolidating and curating data from multiple sources, our dataset allows for an assessment of a broader chemical space and provides a more consistent basis for further analysis. ApisTox encompasses a range of information types, including toxicity levels of four classes of agents: herbicides, fungicides, insecticides and other agrochemicals. All chemical records are additionally augmented with their date of first occurrence in the literature and references to the source chemical database. The data has been rigorously filtered and curated, to ensure accuracy and consistency, addressing a critical need for high-quality, standardized data in bee research.

ApisTox is also one of the very few datasets outside medicinal chemistry that can be used for benchmarking molecular property prediction methods. Currently, machine learning (ML) models are almost exclusively evaluated using data derived from the medicinal fields, whereas agrochemical compounds possess quite different structural and physicochemical characteristics. Our dataset is large enough for training and testing more data-demanding models, being larger than e.g. 17 datasets from Therapeutics Data Commons (TDC) benchmark \cite{TDC}. We also provide explicit train-test splits, including maximum diversity (MaxMin) split and a novel approximation of time split using PubChem literature data. This ensures realistic, challenging evaluation and fair comparison of algorithms. Therefore, ApisTox can serve as a part of benchmarks for novel classification algorithms, including out-of-distribution testing of models designed using medicinal chemistry data.

Finally, the significance of this dataset extends beyond academic research. The dataset can help screen for bee-friendly chemicals and natural products, which help with the development of agricultural systems that promote bee health \cite{bee_health, Chen2022}. It can also be useful for policy decisions and strategies, particularly for evidence based decisions, e.g. assessing the impact of changing regulations and R\&D funding on the number of first-in-class type of compounds developed over time. ApisTox is designed with accessibility and interoperability in mind. This is intended to foster collaboration among researchers and encourage the development of novel, innovative solutions to protect bee populations. In conclusion, the creation of this new, high quality dataset for bee toxicity addresses a critical gap in existing data resources, offering a comprehensive, consistent, and accessible tool for studying the complex factors affecting bee health.

\section*{Methods}
\label{section_methods}

We gather data about the acute honey bee toxicity of pesticides. Following the US EPA guidelines \cite{EPA_guidelines}, as well as EU and UK regulatory standards \cite{PPDB_support_info,EFSA_guidelines}, we focus on the median lethal dose ($LD_{50}$) as a toxicity measure. The same guidelines designate honey bees as a major organism for regulatory decisions, due to their ecological and economic importance.

We enrich it with additional metadata, which allows deeper analyses. Dataset is assembled from various sources, with extensive preprocessing, cleaning and deduplication. The resulting files are described in the \q{Data Records} section, with their characteristics and quality analyzed in the \q{Technical Validation} section.

\subsection*{Data sources}

We base our dataset on three pre-cleaned, high quality data sources, widely utilized in environmental science \cite{PaperUsingEcotox1,PaperUsingEcotox2,BeeToxAI}. They vary in their detail, format and provided metadata.

\textbf{ECOTOX} \cite{ECOTOX} (\href{https://cfpub.epa.gov/ecotox/}{https://cfpub.epa.gov/ecotox/}) database is maintained by US EPA (United States Environmental Protection Agency) and consists of relatively raw data and experimental measurements. For each substance, it typically has many entries from different sources, with varied measurement values, and is the most comprehensive data source for ecotoxicology data. However, it contains relatively less structured data than some other databases, and in particular does not provide SMILES strings, typically processed by computational methods. Instead, it relies on CAS (Chemical Abstracts Service) registry numbers. It is updated quarterly, and we use the version updated on 14th December 2023, which covers almost $1.2$ million measurements and over $13$ thousand chemicals.

\textbf{Pesticide Properties DataBase (PPDB)} \cite{PPDB} (\href{https://sitem.herts.ac.uk/aeru/ppdb/en/index.htm}{https://sitem.herts.ac.uk/aeru/ppdb/en/index.htm}) is a database that catalogs defined pesticide chemical entities, i.e. active ingredients, along with their physicochemical properties, ecotoxicological data, environmental fate, and human health impacts. It was created and is maintained by the Agriculture \& Environment Research Unit (AERU) at the University of Hertfordshire. PPDB is a crucial source of curated, structured data for analyzing pesticide applications and facilitating risk management.

\textbf{Bio-Pesticides DataBase (BPDB)} (\href{https://sitem.herts.ac.uk/aeru/bpdb/index.htm}{https://sitem.herts.ac.uk/aeru/bpdb/index.htm}), also curated by the University of Hertfordshire, is our third source. It provides detailed information on the properties, efficacy, and application of pesticides originating from natural sources, such as microorganisms, plant extracts, and pheromones. BPDB entries often comprise multiple components and lack a defined active ingredient.

Both the PPDB and BPDB are reviewed, managed, and updated through literature and legal resources, following established data curation protocols and guidelines. Those databases received endorsements from leading chemical and agrochemical organizations, including the International Union of Pure and Applied Chemistry (IUPAC) and the Food and Agriculture Organization (FAO). They are continuously updated, and we use data obtained at 22nd February 2024. Raw data obtained at this date is included with other files for reproducibility, as described in detail in section \q{PPDB and BPDB}.

\begin{figure}[hp!]
\centering
\includegraphics[height=0.95\textheight]{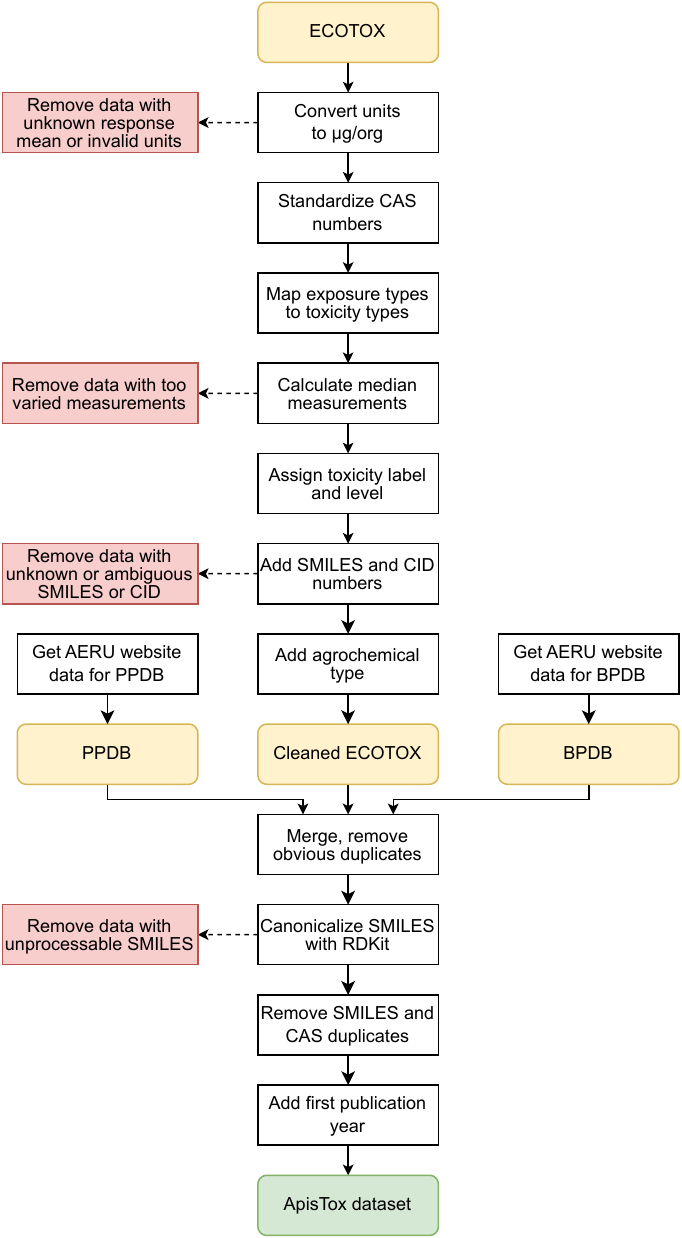}
\caption{Data processing workflow}
\label{fig:workflow_diagram}
\end{figure}

All datasets undergo data processing and cleaning procedures, in order to unify their structure and fill missing fields. The entire workflow has been summarized in Figure \ref{fig:workflow_diagram}. The workflow itself constitutes a major novel contribution in this work, fully describing the set of procedures for parsing, merging, cleaning and deduplication of those three databases, with the reusable, open source code available.

During the entire data cleaning and processing, we save all removed rows, along with removal reason, in a separate file, and include it along the rest of the raw and processed data (see \q{Data Records} section for details). Manual analysis of this data allowed us to verify the correctness of processing steps. In further sections, we describe them, and their rationale, in detail.

\subsection*{Classification labels}

For all three databases, we apply the same procedure for creating the classification labels of honey bee toxicity, based on $LD_{50}$ values. As such, ApisTox can be used for classification, rather than regression, and here we explain the rationale of this decision.

Toxicity studies, in particular QSAR models, often use classification approach, rather than regression. Examples include Tox21 and ToxCast datasets from MoleculeNet \cite{MoleculeNet}, as well as most toxicity datasets from TDC \cite{TDC}. If direct measurement values are available, most often $LD_{50}$ or $LC_{50}$, they can be binarized into toxic/non-toxic classes, using some predefined threshold. Sometimes more classes are used, e.g. ternary toxic / moderately toxic / highly toxic. Such approach has also been used e.g. for rats \cite{ld50_rats,ld50_rats_2}, fish and \textit{Daphnia Magna} \cite{ld50_fish}, and earthworms \cite{ld50_earthworms}.

Such approach is used particularly due to the inherent uncertainty of toxicity measurements, in particular for $LD_{50}$. It is a statistically obtained virtual value, constituting a best-effort estimation of the dose required to kill 50\% of animals \cite{ld50_toxicology}. There are many different protocols for calculating $LD_{50}$, differing in complexity and quality \cite{ld50_toxicology,ld50_measurements}. For experiments on rats, $LD_{50}$ values can vary as much as 3- to 11-fold between different laboratories \cite{ld50_measurements_rats}, and similarly significant differences have been observed for honey bees \cite{ld50_apis}. For low toxicity pesticides (with $LD_{50}$ values greater or equal to $100$ $\mu \text{g / bee}$), measuring $LD_{50}$ for honey bees is particularly problematic, and EFSA guidelines \cite{EFSA_guidelines} contain dedicated procedures and limit tests for those cases. As such, the mean or median aggregate value for regression would itself be highly uncertain. This is also why PPDB and BPDB often don't provide exact $LD_{50}$ values, but rather a range limit, e.g. \q{>300}. In classification, we only require the estimated $LD_{50}$ values to lie below or above a given threshold, lessening the impact of measurement errors.

Furthermore, in ecotoxicology regulatory organs recognize this issue and provide the official $LD_{50}$ thresholds for different organisms. In practice, when designing new pesticides, one is not necessarily interested in the exact $LD_{50}$ value but rather if the new substance will be accepted by the regulatory organ, i.e. if the toxicity is low enough. Classification models explicitly provide this information, when an appropriate threshold is used, with predicted probability giving a measure of certainty in the prediction.

US EPA guidelines \cite{EPA_guidelines} for honey bees use binary toxic/non-toxic labels, with $11$ $\mu \text{g / org}$ as the threshold for acute bee toxicity. This means that $LD_{50}$ less or equal to this value is deemed toxic to honey bees. We mark toxic class as $1$, and non-toxic as $0$, as shown in Equation \ref{eq:1}.

\begin{equation} \label{eq:1}
label(LD_{50}) = 
    \begin{cases}
      1 & \text{if}\quad LD_{50} \leq 11 \\
      0 & \text{if}\quad LD_{50} > 11
    \end{cases}       
\end{equation}

PPDB, based on EU and UK regulatory guidelines \cite{PPDB_support_info,EFSA_guidelines}, introduced a 3-level (ternary) highly toxic/moderately toxic/non-toxic labels, with thresholds $1$ $\mu \text{g / org}$ and $100$ $\mu \text{g / org}$. We mark highly toxic as $2$, moderately toxic as $1$, and non-toxic as $0$, as shown in Equation \ref{eq:2}.

\begin{equation} \label{eq:2}
ppdb\_level(LD_{50}) = 
    \begin{cases}
      2 & \text{if}\quad LD_{50} \leq 1 \\
      1 & \text{if}\quad 1 < LD_{50} \leq 100 \\
      0 & \text{if}\quad LD_{50} > 100
    \end{cases}       
\end{equation}

To maximize the applicability of ApisTox, we calculate classes using both of those approaches: binary \textit{label} and ternary \textit{ppdb\_level}. As such, it can be used either for binary classification or ordinal classification with three levels.

\subsection*{ECOTOX processing}

ECOTOX database offers a rich set of filters, allowing flexible retrieval of data. It is divided into \q{Aquatic} and \q{Terrestial} organisms. We select the latter, and apply the following filters:
\begin{enumerate}
\item Effect \q{mortality} - effects are observed responses of different types when applying a given chemical to selected species. Mortality is the most common target for toxicological studies.
\item Endpoint \q{LD50} - endpoints are measurable numerical targets for given effects. We use median lethal dose ($LD_{50}$), for consistency with PPDB and BPDB, following EPA guidelines \cite{EPA_guidelines} and EU and UK regulatory standards \cite{PPDB_support_info, EFSA_guidelines}.
\item Species contains \q{apis mellifera} - we select \textit{Apis mellifera} (honey bee) data. We use contains filter 
instead of exact match, because while the vast majority of measurements use the general species name, some also specify the subspecies, most commonly \textit{Apis mellifera mellifera} (European dark bee).
\item Test locations \q{lab} - we exclude less reliable field measurements and measurements without stated test location. This filters out 10 rows.
\end{enumerate}

We do not filter data by observed duration, because it contains measurements with at most 8 days, which still corresponds to acute toxicity, and most measurements are for 72 hours or less. The resulting set contains 2674 raw measurements, which then undergo a series of processing steps to obtain data aggregated per pesticide. For each measurement, we select substance name, CAS number, exposure type, observed response mean and unit. They identify the pesticide used, how it was applied, and measured $LD_{50}$.

Firstly, we remove all rows without mean response value, marked as \q{NR} (Not Reported). Then, we standardize measurements to unit $\mu \text{g / bee}$ (also known as $\mu \text{g / org}$). This unit is used by PPDB, BPDB, and US EPA guidelines for measuring bee toxicity \cite{EPA_guidelines}. Overall, there are over 40 different units used. We convert those that can be unambiguously mapped to $\mu \text{g / bee}$ (e.g. \q{pg/org}, \q{ug/bee}), and remove all other samples. Removed units are, e.g. \q{\%}, \q{ae ug/org:x} or \q{mg/cm2}.

ECOTOX uses non-standard notation for CAS numbers, without parentheses. We normalize such cases, e.g. changing 1194656 to {1194-65-6}.

Toxicity is typically divided into oral, contact, or other way of application. PPDB and BPDB also use this system. However, ECOTOX applies a more fine-grained classification, so we map the ECOTOX exposure types into three standardized toxicity types as follows:
\begin{itemize}
    \item \q{Diet, unspecified}, \q{Drinking water}, \q{Food} - oral
    \item \q{Dermal}, \q{Direct application}, \q{Topical, general} - contact
    \item \q{Multiple routes between application groups}, \q{Oral via capsule}\footnote{Only a single row had the \q{Oral via capsule} value, and this was not a standard oral diet in the original experimental study\cite{oral_via_capsule}, so we include it as \q{other}, rather than \q{oral}.}, \q{Spray, unspecified}, \q{Environmental, unspecified} - other
\end{itemize}

To obtain a dataset with a single pesticide per row, raw $LD_{50}$ measurements from ECOTOX need to be aggregated into a single number. We identify each distinct substance by the CAS number. They can have $LD_{50}$ measurements for oral, contact, and other toxicity type. For many pesticides, only one or two of those are available. Their values vary strongly for some pesticides, so we take a conservative approach here, to ensure high data quality.

For each toxicity type, we check if the lowest and highest measurement agree for the EPA guidelines threshold, i.e. if all values are below or above $11$ $\mu \text{g / org}$. If so, we assign toxic or non-toxic class, respectively. Otherwise, we mark the toxicity label as \q{Unspecified}, and remove such rows. This way, we further alleviate problems arising from $LD_{50}$ measurement uncertainty, as described in the \q{Classification labels} section.

We also assign PPDB ternary level based on median measurement value. Lastly, for each pesticide, we take the strongest toxicity type, as representing the most toxic effect the pesticide can have on bees. For example, if it has oral toxicity $2$ (highly toxic) and contact toxicity $1$ (moderately toxic), we mark it as $2$. This way, only pesticides truly safe for honey bees will get level $0$. We also save the information which toxicity type was the strongest for each pesticide.

Further, we add SMILES strings and PubChem CID (Compound ID) numbers. SMILES strings are required by computational libraries, and CID numbers enable easy and unambiguous lookup of molecules in PubChem \cite{PubChem}, the largest openly available database of chemical information. We utilize PUG REST API \cite{PubChem_PUG_REST}, mapping CAS numbers to SMILES strings and CID numbers. We remove those molecules for which that operation was impossible or ambiguous.

Lastly, we add information about agrochemical type for each substance, i.e. whether it is a herbicide, fungicide, insecticide, or other agrochemical. This allows users to carry out more detailed analyses, e.g. narrow down toxicity studies only to insectides, which are typically the most dangerous pesticide type for bees. In addition, predictive models and their performance can be analyzed for those types. For example, one can check if the model performs equally well for fungicides and insecticides.

In PubChem, compounds can have a separate section \q{Agrochemical information} with descriptions of agricultural applications. Using PUG REST API, we fetch descriptions from this section for each substance, and we search for keywords \q{herbicide}, \q{fungicide} and \q{insecticide}. If any of those words appear in the description, we note that information as agrochemical type. If none are found, but the compound has this section, it means that it has other agrochemical applications, e.g. as a growth agent or fertilizer, and we mark it as \q{other agrochemical}. This creates a total of four boolean (binary) variables. If the compound page does not have \q{Agrochemical information} section, all four type features are zeros, meaning \q{unknown} pesticide type. We still include those substances, since we have bee toxicity measurements for them. They could have been researched as potential pesticides, but not yet registered or widely used in agriculture.

\subsection*{PPDB and BPDB}

PPDB and BPDB databases have similar structure, organized in HTML tables, available via web pages. They contain a very comprehensive set of information about pesticides (over 300 different fields). We extract only some fields, from which we compute relevant features using regular expressions on HTML responses, as described below. All molecules that do not have toxicity measurements for honey bees are ignored. This approach always utilizes the latest available data, and we use the data obtained on 22nd February 2024.

For each pesticide, $LD_{50}$ values for oral, contact, and other honey bee toxicity can be provided. If we have more than one toxicity type, we take the lowest value, i.e. the strongest toxicity. We create binary label and ternary level, as described in the \q{Classification labels} section, similarly to ECOTOX processing. Additionally, some values are provided in a non-standard format, especially for low toxicity, and marked as \q{Low}, \q{Non-toxic}, or in scientific notation, e.g. \q{10\textasciicircum 3}. We normalize those cases as non-toxic, i.e. label and level 0.

We combine \q{Summary}, \q{Description} and \q{Pesticide type} fields into a single text to extract the pesticide type. We search for keywords \q{herbicide}, \q{fungicide} and \q{insecticide}, similarly to processing PubChem data for ECOTOX. Since PPDB and BPDB databases contain only agrochemicals, if none of those words appear, we mark a given substance as \q{other agrochemical}.

CAS numbers and SMILES strings are available for almost all pesticides, and we exclude those without this information. CID numbers are also often available, but in case they are missing, we fill them using PubChem and PUG REST API, based on CAS and SMILES.

\subsection*{Molecular standardization and deduplication}

After gathering three preprocessed datasets, we merge them. Firstly, we concatenate all rows and drop obvious duplicates, i.e. compounds with the same CAS, SMILES and label. However, since SMILES strings are not unequivocal, this does not yet fully ensure lack of structural duplicates among molecules. Thus, we standardize molecules with RDKit and create canonical SMILES, which can be used for more reliable deduplication.

We perform multi-step canonicalization and sanitization based on input SMILES, using RDKit 2023.9.5 \cite{RDKit}. This way, we make sure that all SMILES strings are valid and processable by this software, and all molecules conform to basic sanity checks like proper valence, all bonds are represented in a unified way etc. This is done with \texttt{MolFromSmiles} function, which calls a whole sanitization pipeline underneath \cite{RDKit_sanitization}, and with \texttt{MolToSmiles} function, which generates canonical SMILES. This pipeline includes:
\begin{itemize}
    \item valence cleanup, handling non-standard valence states
    \item kekulization
    \item determining aromaticity of bonds and atoms
    \item calculation of bond conjugation and atom hybridization states
    \item chirality cleanup, removing chiral tags from atoms that are not sp3 hybridized
    \item adjusting hydrogens, leaving only necessary explicit ones (e.g. for heteroatoms in aromatic rings)
\end{itemize}

Those steps are common for most molecular standardization workflows \cite{standardization_1,standardization_2,standardization_3}. We perform quite conservative set of operations, to allow end users to use their own more aggressive transformations or filtering if desired. In particular, following EPA guidelines for pesticide QSAR (created in conjunction with NAFTA Technical Working Group on Pesticides) \cite{EPA_QSAR_guidelines}, we do not remove salts, inorganics, mixtures, or metal-containing compounds. This is particularly relevant for toxicity studies of organic salts \cite{standardization_pesticides}. Most standardization workflows focus solely on medicinal chemistry, where such molecules are commonly removed, but this would get rid of much important information prevalent in agrochemistry. For example, Mancozeb and Maneb are treated as independent pesticides in regulatory issues \cite{mancozeb_maneb}, while they only differ in the complexation with metal ions (Mancozeb is complexed with manganese and zinc ions, Maneb is complexed only with manganese ions).

Such standardization ensures that all agrochemicals in ApisTox are processable by RDKit, which is the most commonly used open source chemoinformatics tool in Python, and also integrates with e.g. KNIME. Resulting SMILES strings are canonical, and can be used to perform further deduplication. We then drop all duplicates, first using SMILES, and then using CAS. When removing duplicates, we keep rows from datasets in the order of preference: PPDB, BPDB, ECOTOX. This was motivated by the fact that PPDB and BPDB are additionally manually verified, and therefore can have slightly higher quality.

\subsection*{Adding PubChem literature date}

Lastly, we add the first publication year information to all molecules, using their CID numbers and PubChem literature records. While this is not exact information about first usage of a given substance as a pesticide, it provides a good approximation of when a given compound has entered the public domain. With this information, additional analyses can be carried out with ApisTox, related to the development of pesticide structures through time and broader analysis of the effectiveness of regulations and R\&D spendings. It can also be directly used in molecular property prediction to create a time split (also known as chronological split) for train-test sets. This approximation should be much more precise than scaffold split \cite{MoleculeNet} or SIMPD \cite{SIMPD}. Those techniques were created to approximate time split, but are based on some predefined structural biases, e.g. Bemis-Murcko scaffolds, which PubChem dates do not enforce.

We noticed that PubChem has three obvious errors in this regard, attributing an unreasonably large number of molecules to three publications \cite{PubChem_citation_err_1,PubChem_citation_err_2,PubChem_citation_err_3}. We verified manually that this is a mistake, and we exclude those papers from this mapping. We sort the resulting dataset by this year. It contains a total number of 1035 molecules.

\subsection*{Data splitting}

Statistical models, especially predictive machine learning (ML) models, require separate parts of data for fitting (training) models, and for their testing and verification of performance. For molecular graphs and chemical data especially, there are many non-obvious possible sources of data leakage, which lead to improper and overly optimistic estimation of models' performance. Having predetermined splits is beneficial for reproducible science in ML, therefore we compute train-test splits and distribute those files along the full dataset. To the best of our knowledge, ApisTox is the first dataset of honey bee toxicity to provide explicit train-test splits and evaluation protocol, making it better suited for comparison of ML techniques.

We split the dataset in three different ways: stratified random split, time split, and MaxMin split. Random and MaxMin are interpolative, validating predictive performance inside the domain of the dataset. Time split is extrapolative, meaning that it aims to verify the out-of-domain generalization of models to structurally novel compounds. In all cases, we use 80\%-20\% proportions, resulting in 828 training and 207 testing molecules.

We do not provide scaffold split \cite{MoleculeNet}, popular in medicinal chemistry, for the following reasons. It uses the Bemis-Murcko scaffolds to group molecules, and then puts the smallest groups in the test set. The idea is to put the most structurally different compounds into the test set, which requires generalization to new areas of chemical space. This way, it aims to approximate a time split, with an assumption that future compounds have new scaffolds. However, this assumes that those scaffolds are structurally meaningful, and that the dataset will contain many small groups of scaffolds. Bemis-Murcko scaffolds are not defined for ring-free compounds, or for those with disconnected components (e.g. salts), since their definition relies on connected ring systems \cite{Bemis_Murcko}. Such molecules constitute almost 20\% of the data (see the \q{Technical Validation} section for more details), and that group would result in one large \q{no scaffold} group in the training set. Therefore, this approach does not differentiate them structurally and, in the worst case, almost identical molecules can be in training and test sets, introducing data leakage. This is clearly a problem, and for this reason we recommend using MaxMin or time split instead. See also \cite{scaffold_problems} for other potential caveats with scaffold-based approaches.

\textbf{Stratified random split} puts data points randomly into training and testing parts, disregarding the features or internal structure of molecular graphs. Stratification ensures that the proportion of toxic and non-toxic classes is approximately the same in the full dataset and both splits. This is desirable, since ApisTox is imbalanced and toxic molecules constitute a minority class. We use binary toxicity labels here. Such split is susceptible to clustering in chemical space, and therefore can result in structurally similar pesticides in both training and testing sets \cite{splitting_data,MoleculeNet}. Depending on application, this may be seen as a form of data leakage and artificially increase test score.

\textbf{Time split} puts the newest molecules in the testing set, in order to simulate the actual discovery and adoption of pesticides. The underlying assumption is that intrinsically new substances are introduced over time. This is often a very realistic setting, especially for designing novel molecules, but such information is rarely available in molecular datasets \cite{splitting_data,split_time}. Our literature-based year assignment, while not perfectly precise, allows using this kind of split. Using PubChem to approximate the time split is, as far as we are aware, a novel approach, and a meaningful alternative to e.g. scaffold split or SIMPD method \cite{SIMPD}.

\textbf{MaxMin split} utilizes the maximum diversity picking algorithm to select test molecules such that the sum of distances in the test set is maximized \cite{split_maxmin,split_maxmin_2}. Typically, ECFP4 fingerprints with either Tanimoto or Dice distance are used for this purpose. This way, the test set has very high coverage of the chemical space and tests the generalization of the algorithm to all kinds of compounds in the data. Due to maximization of test set distances, it selects molecules much more uniformly in the chemical space than random split, alleviating the problem of clustering.

We use Scikit-learn \cite{scikit-learn} for computing stratified random split, Pandas \cite{Pandas} for time split, and DeepChem \cite{DeepChem} for MaxMin split (with default Dice distance and 1024 bits ECFP4 fingerprints).

\section*{Data Records}

ApisTox dataset is available on Zenodo \cite{ApisTox_Zenodo}, as well as on GitHub \cite{ApisTox_GitHub}. In both places, we distribute the files on all stages of processing: raw, cleaned, the final dataset, and data splits as described in the \q{Methods} section. All files are in human-readable CSV format. The main dataset file is \textit{dataset\_final.csv} (in \textit{outputs} directory on GitHub), with columns described in Table \ref{tab:dataset_columns}.

\textit{ecotox.csv} contains raw outputs of ECOTOX database query, and uses pipe \q{|} as separator. All other files use commas as separators. \textit{ppdb.csv} and \textit{bpdb.csv} files consist of data from PPDB and BPDB, respectively. For those two databases, we distribute only the subset of molecules that contained any bee pesticide data: 860 from PPDB and 122 from BPDB. From raw data available on the website, we only include the public identifiers: name, CID, CAS, SMILES. Columns \q{herbicide}, \q{fungicide}, \q{insecticide}, \q{other\_agrochemical}, \q{label} and \q{ppdb\_level} are features created from HTML responses, as described in the \q{Methods} section. On GitHub, those files are in the \textit{raw\_data} directory.

\textit{ecotox\_cleaned\_data.csv} file contains ECOTOX data after processing and cleaning, and \textit{excluded\_data.csv} contains rows removed at any point of processing, along with reason for exclusion. \textit{dataset\_final.csv} is the main dataset file. On GitHub, those files are in the \textit{outputs} directory.

For data splits, described in the \q{Methods} section, we provide training and testing subset of data after splitting, in files \textit{random\_train.csv}, \textit{random\_test.csv}, \textit{time\_train.csv}, and \textit{time\_test.csv}, \textit{maxmin\_train.csv}, \textit{maxmin\_text.csv}. Those files have exactly the same structure as the main dataset file. On GitHub, they are in \textit{outputs/splits} directory.

\begin{table}[h]
\centering
\begin{tabular}{|c|c|c|}
\hline
\textbf{Column}     & \textbf{Type} & \textbf{Description}                                      \\ \hline
name                & string        & Chemical name                                             \\ \hline
CID                 & integer       & PubChem Compound ID number                                \\ \hline
CAS                 & string        & Chemical Abstracts Service registry number                \\ \hline
SMILES              & string        & Molecule structure in SMILES format                       \\ \hline
source              & string        & Compound source: ECOTOX, PPDB or BPDB                     \\ \hline
year                & integer       & First publication year in literature according to PubChem \\ \hline
toxicity\_type      & string        & Strongest toxicity type: Contact, Oral or Other           \\ \hline
herbicide           & boolean       & Is the chemical used as a herbicide?                      \\ \hline
fungicide           & boolean       & Is the chemical used as a fungicide?                      \\ \hline
insecticide         & boolean       & Is the chemical used as an insecticide?                   \\ \hline
other\_agrochemical & boolean       & Is the chemical used in other way as an agrochemical?     \\ \hline
label               & boolean       & Binary toxicity label                                     \\ \hline
ppdb\_level         & integer       & Ternary toxicity level                                    \\ \hline
\end{tabular}
\caption{Features in the final ApisTox dataset.}
\label{tab:dataset_columns}
\end{table}

\section*{Technical Validation}

In this section, we present molecular characteristics of ApisTox dataset, and technical analyses of basic properties relevant to chemoinformatical applications. We use the data from \textit{dataset\_final.csv} file. Suggestions for further modeling and applications are in the \q{Usage Notes} section.

\subsection*{Dataset quality verification}

Here, we perform the basic quality checks for molecular data. ApisTox, as a curated and unified collection of data, should cover all three source databases. We compare the number of molecules of the final dataset and source databases in Table \ref{tab:dataset_sources_sizes}, all after the same data cleaning procedure outlined in the \q{Methods} section. ApisTox is indeed larger than all input datasets, validating our data combination process. In particular, it is also almost 25\% larger than PPDB, the largest of the source datasets.

\begin{table}[H]
\centering
\begin{tabular}{lcccc}
\toprule
& \textbf{ApisTox} & \textbf{PPDB} & \textbf{BPBD} & \textbf{ECOTOX} \\
\midrule
Total molecules & 1035  & 831   & 115   & 521 \\
Toxic molecules & 296   & 228   & 18    & 189 \\
Non-toxic molecules & 739   & 603   & 97    & 332 \\
\bottomrule
\end{tabular}
\caption{Comparison of the number of molecules between ApisTox and the source databases.}
\label{tab:dataset_sources_sizes}
 \end{table}

\begin{figure}[ht]
\centering
\includegraphics[width=0.5\textwidth]{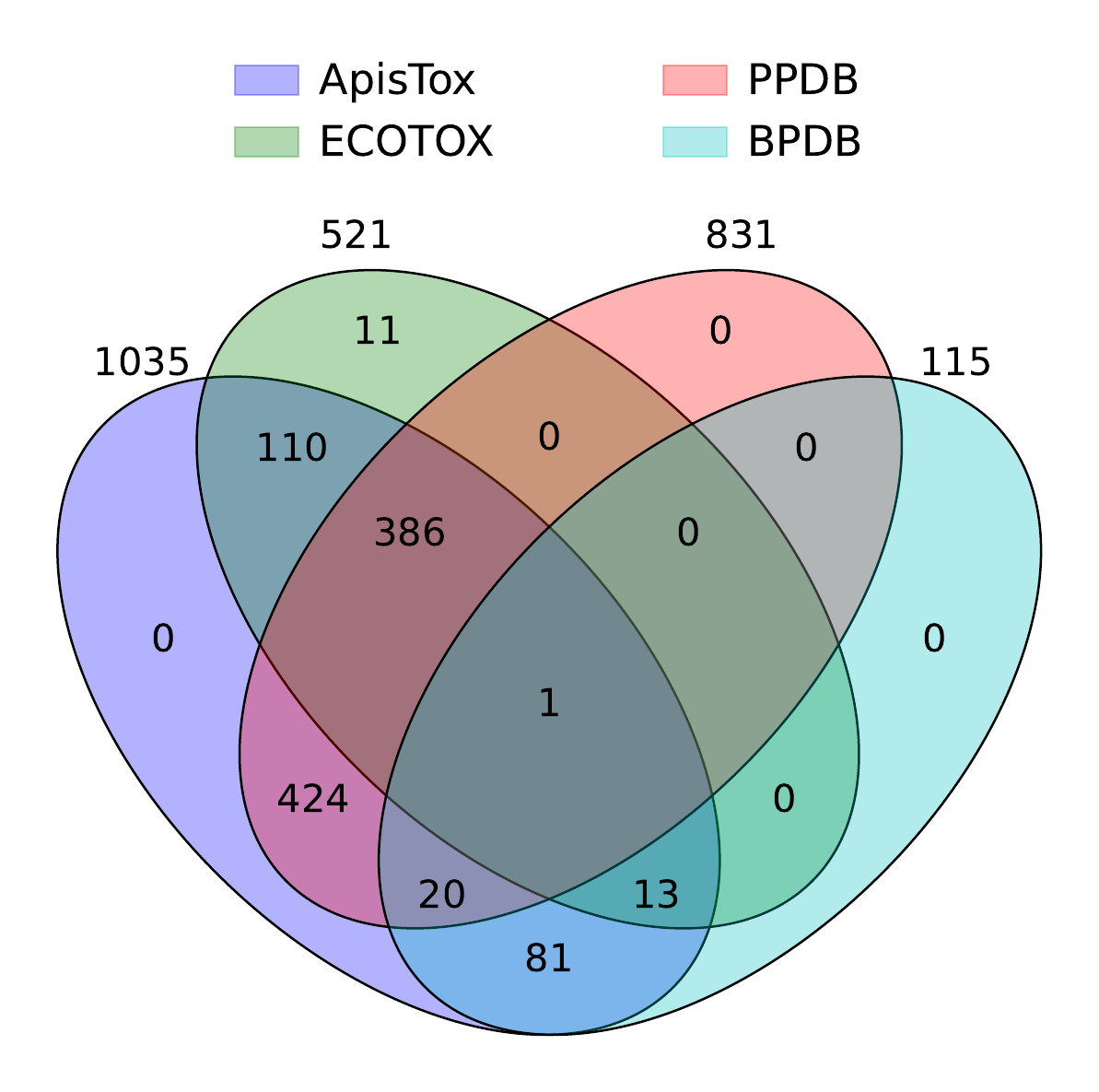}
\caption{Number of molecules present in ApisTox sourced from three databases.}
\label{fig:venn}
\end{figure}

Furthermore, we verify that all canonical SMILES from the cleaned source datasets are included in the final dataset. Results are summarized by a Venn diagram in Figure \ref{fig:venn}. We integrated molecules from each data source, creating a significantly larger and unified dataset beneficial for ecotoxicology research. The final dataset comprises all filtered molecules from sourced databases, except for 11 SMILES strings from ECOTOX. Manual checks showed that those molecules are subtle duplicates, having the same CAS numbers as molecules from PPDB already included. This is the consequence of non-uniqueness of SMILES format, i.e. a given molecule can be written in many ways, and even RDKit standardization and canonicalization workflow is not always able to detect such cases. This shows that a two-step deduplication procedure, which includes both canonical SMILES and CAS deduplication, is indeed required for proper merging of molecular datasets (see the \q{Methods} section for details).

We also verify that ApisTox contains only valid entries that can be processed by RDKit, and contains no duplicated molecules (in terms of canonical SMILES and CAS numbers). To further validate our preprocessing workflow in this regard, we apply it to datasets proposed for honey bee toxicity analyses: CropCSM \cite{CropCSM}, BeeTOX \cite{BeeTox} and BeeToxAI \cite{BeeToxAI}. As shown in Table \ref{tab:datasets_quality_comparison}, other datasets contain invalid SMILES and duplicates, even as much as 36\% for BeeToxAI. This emphasizes the need for highly rigorous data processing workflows like ours. ApisTox is considerably larger than all of them, containing many more toxic molecules in particular, which are crucial for understanding the underlying causes of pesticide toxicity for honey bees.

\begin{table}[H]
\centering
\begin{tabular}{lcccc}
\toprule
& \textbf{ApisTox} & \textbf{CropCSM} & \textbf{BeeTOX} & \textbf{BeeToxAI} \\
\midrule
Initial number of molecules & 1035  & 900   & 891   & 734 \\
Invalid entries & 0   & 1   & 3    & 0 \\
Duplicated molecules & 0   & 28   & 12    & 262 \\
Cleaned dataset size & 1035   & 871   & 876    & 472 \\
Non-toxic molecules & 739   & 638   & 645    & 282 \\
Toxic molecules & 296   & 233   & 231    & 121 \\
\bottomrule
\end{tabular}
\caption{Comparison of the ApisTox and previous datasets on pesticide toxicity for honey bees. In case of BeetoxAI, only 403 molecules were reported with toxicity labels.}
\label{tab:datasets_quality_comparison}
\end{table}

\subsection*{Molecular properties}

\begin{figure}[ht]
\centering
\includegraphics[width=\textwidth]{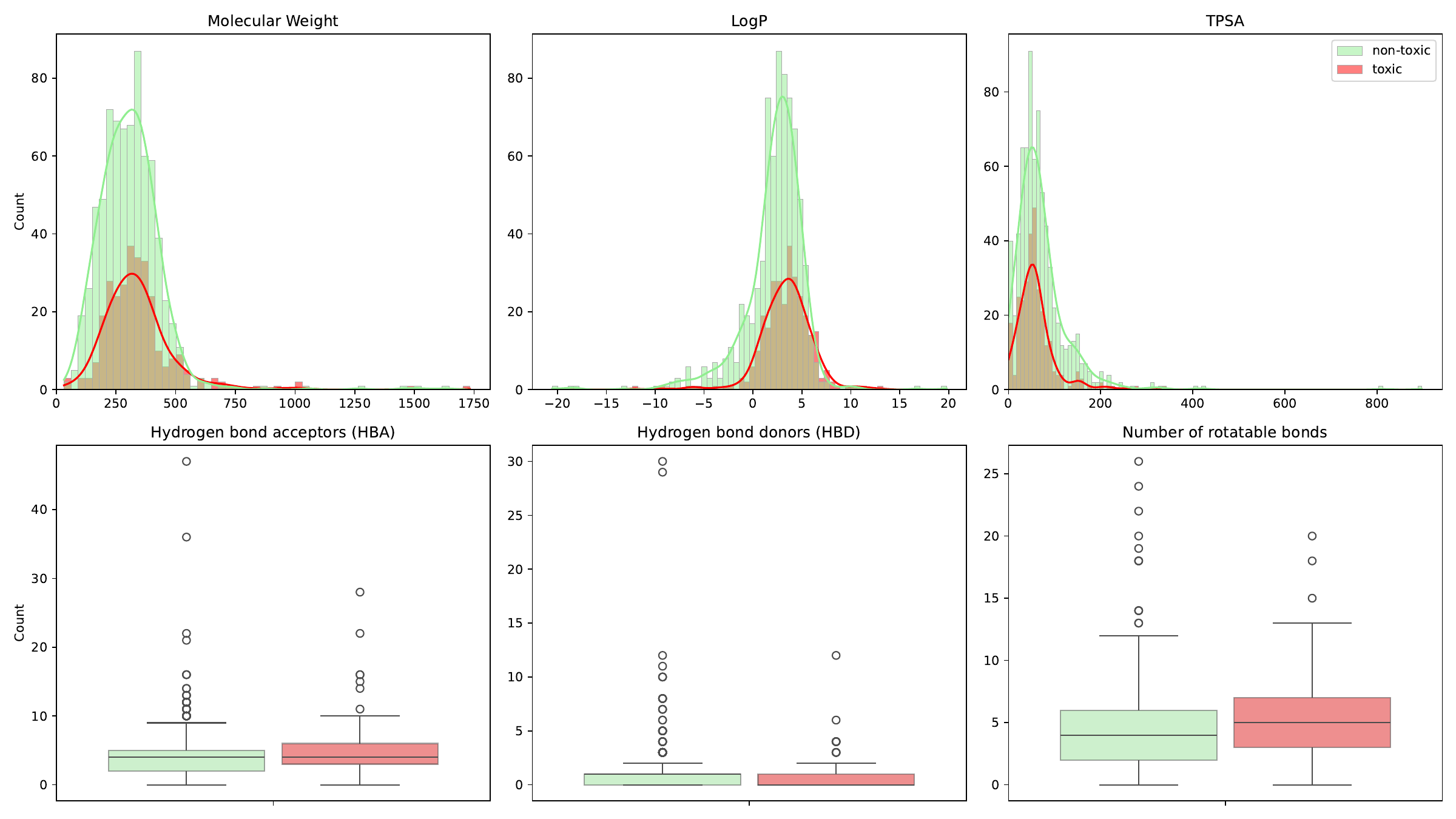}
\caption{Comparison of basic physico-chemical properties of molecules in ApisTox for non-toxic and toxic pesticides.}
\label{fig:physicochemical_properties}
\end{figure}

We present distributions of six basic physico-chemical molecular properties in Figure \ref{fig:physicochemical_properties}: molecular weight (MW), logarithm of octanol-water partition coefficient (logP), topological polar surface area (TPSA), number of hydrogen bond acceptors and donors (HBA, HBD) and number of rotatable bonds.

Distributions for non-toxic and toxic pesticides are similarly shaped. Positive logP and low TPSA (<100) are dominant, suggesting that most molecules are non-polar. This makes sense for pesticides, since non-polar molecules penetrate biological membranes much more effectively than polar ones \cite{non_polar_membrane_penetration}. Toxic molecules have slightly higher HBA and number of rotatable bonds, across all quartiles. ApisTox contains many large molecules, as measures by molecular weight, with some over 1000 daltons. This also explains outliers in terms of HBA and HBD.

\subsection*{Toxicity labels distributions}

\begin{figure}[ht]
    \centering
    \includegraphics[width=\textwidth]{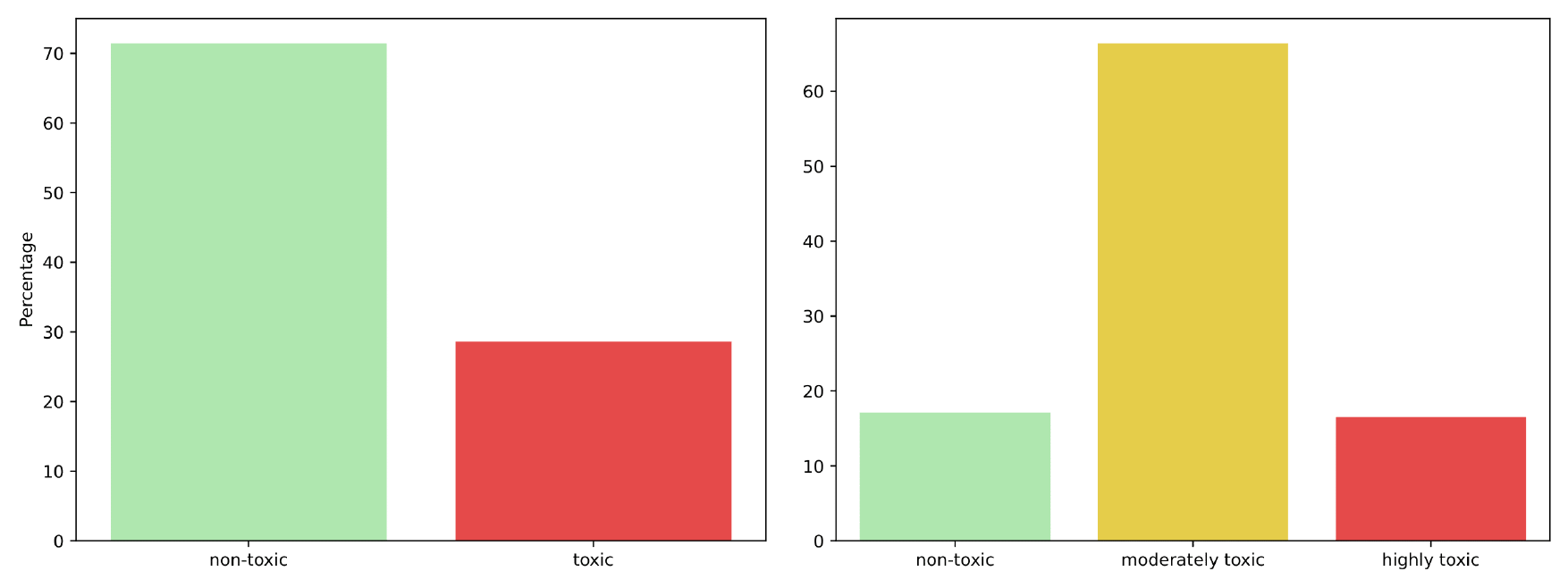}
    \caption{Toxicity binary labels and ternary levels distributions.}
    \label{fig:labels_distributions}
\end{figure}

We present distributions of toxicity labels in Figure \ref{fig:labels_distributions}. In terms of binary toxicity label, ApisTox is moderately imbalanced, with pesticides toxic for honey bees constituting 29\% of the data. Concretely, there are 739 non-toxic and 296 toxic compounds. The ternary level is more severely imbalanced, with 17\% non-toxic, 66\% moderately toxic, and 17\% highly toxic molecules. This is due to the very wide definition of moderate toxicity in PPDB. Under this methodology, we have 177 non-toxic, 687 moderately toxic and 171 highly toxic molecules.

Such imbalance in class distributions influences metrics appropriate for validating predictive models, as discussed in the \q{Usage Notes} section.

\subsection*{Splits analysis}

We analyze the effects of different proposed splits into training and testing sets. Preferably, both parts should have similar distribution of classes and pesticide types, to keep the basic characteristics of the dataset similar and have good representation of all data segments in the test set.

In addition, we want the test set to be diverse, and to be reasonably structurally different from the training data in order to avoid data leakage. For measuring those qualities, we utilize commonly used ECFP4 (Morgan) fingerprints with 1024 bits and Tanimoto distance (one minus Tanimoto similarity), which allow us to represent the molecules in a vector space. We measure diversity by calculating the average distance between test molecules, as high values mean that we avoid \q{clustering} the test samples, which would measure generalization only in small subsets of a chemical space. To calculate structural separation, we compute average distance between each test sample and closest training sample. A high value of such metric means that test molecules are structurally different from training ones, and we avoid data leakage from too similar compounds.

We summarize the class distributions and distances in Table \ref{tab:distributions}. An additional table with pesticide types distributions is available in the supplementary information.

Class distributions for both binary labels and ternary PPDB levels are similar in all cases, which indicates that no splitting method introduces unwanted additional imbalance. The diversity is highest for MaxMin split and the lowest for stratified random split, which follows their motivation outlined in the \q{Methods} section. Train-test separation is also the highest for MaxMin split, meaning that not only its test set covers the chemical space of the dataset well, but also at the same time is not too similar to training compounds.

\begin{table}[H]
\centering
\resizebox{\textwidth}{!}{
\begin{tabular}{|c|cc|cc|c|c|}
\hline
\textbf{}           & \multicolumn{2}{c|}{\textbf{Binary label}}     & \multicolumn{2}{c|}{\textbf{Ternary level}}                  & \textbf{Test set diversity} & \textbf{Train-test separation} \\ \hline
\textbf{Split type} & \multicolumn{1}{c|}{Train}       & Test        & \multicolumn{1}{c|}{Train}              & Test               & -                           & -                              \\ \hline
Stratified random   & \multicolumn{1}{c|}{71\% / 29\%} & 71\% / 29\% & \multicolumn{1}{c|}{65\% / 18\% / 17\%} & 70\% / 16\% / 14\% & 0.471                       & 0.900                          \\ \hline
Time                & \multicolumn{1}{c|}{69\% / 31\%} & 80\% / 20\% & \multicolumn{1}{c|}{66\% / 18\% / 16\%} & 69\% / 19\% / 12\% & 0.515                       & 0.882                          \\ \hline
MaxMin              & \multicolumn{1}{c|}{69\% / 31\%} & 80\% / 20\% & \multicolumn{1}{c|}{66\% / 18\% / 16\%} & 67\% / 22\% / 11\% & 0.611                       & 0.944                          \\ \hline
\end{tabular}
}
\caption{Dataset splits statistics.}
\label{tab:distributions}
\end{table}

\subsection*{Pesticides timeline}

\begin{figure}[ht]
\centering
\includegraphics[width=0.75\textwidth]{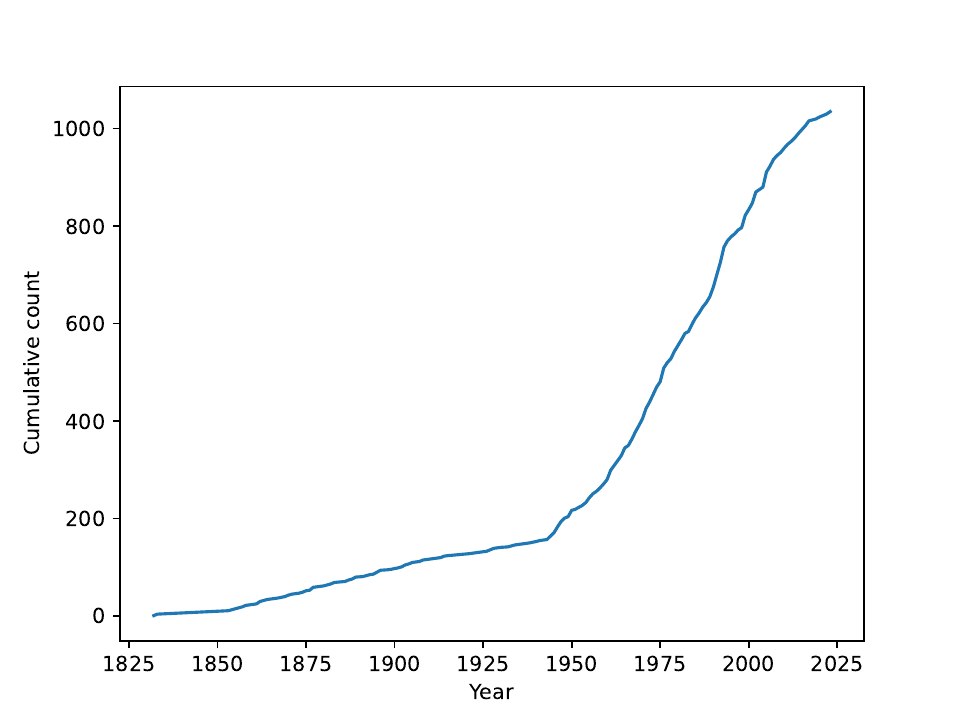}
\caption{Cumulative count of pesticides by year.}
\label{fig:timeline}
\end{figure}

Using literature publication dates from PubChem, we present a timeline plot with the total number of available pesticides per year in Figure \ref{fig:timeline}. The results align with agrochemical literature, with the oldest pesticides like benzoic acid or calcium carbonate known and used in the 19th century, and the majority of older generation pesticides (often toxic and outdated by contemporary standards) developed in the 1970s (carbamates), 1980s (pyrethroids) and 1990s (neonicotinoids) \cite{pesticides_impact_history,pesticides_neonicotinoids}. The decrease in new developments in the 21st century is also supported by both literature and industry trends, e.g. no herbicides with new mode of action have been introduced commercially in the last 30 years \cite{pesticides_21st_century}.

\subsection*{Molecular filter rules}

A common approach to drug design in medicinal chemistry is the application of molecular filters \cite{molecular_filters}. They consist of conditions (rules) that have to be satisfied by new drug candidates, ensuring that they are bioavailable, have high absorption and permeation, or have other desirable properties. Filter rules are typically based on statistics of molecule properties derived from large collections of compounds of a particular type, e.g. drug-like molecules. While the filter-based approach is conservative and may limit the diversity of novel compounds for drug design, they can also be used to verify the quality of the data. For high-quality datasets, a reasonable percentage of compounds should meet the requirements of typical filters.

In the context of pesticides, multiple filters have been designed and used. The most widely used Lipinski's rule of five has been designed for bioavailable, drug-like molecules with high absorption and permeation \cite{filters_Lipinski}. It has been shown that it also works for pesticides and their subtypes (e.g. insecticides) \cite{filters_Tice}, which also have to possess similar bioavailability properties. Specialized filters for agrochemistry have been designed, most prominently Hao's pesticide filter \cite{filters_Hao}, and Tice's filters for herbicides and insecticides \cite{filters_Tice}. Since filters are quite conservative and can often reject specific groups of molecules (e.g. macrolides for Lipinski's rule), a common variant allows violating one of the conditions.

We present results of Lipinski, Hao and two Tice filters in Table \ref{tab:filters}, i.e. what percentage of the data fulfill the given filter conditions. We analyze both the whole ApisTox dataset and subsets with particular pesticide types. Results are provided for all rules satisfied, and for the more relaxed variant with one violation allowed.

The vast majority of both the full dataset and pesticide type subsets fulfill those filters when one violation is allowed. This means that our data follows established rules for bioavailable drugs, pesticides, and their types. Even when no violation is allowed, the majority of ApisTox molecules satisfy the filters' conditions. At the same time, lower percentages for Hao and Tice filters with all rules indicate that data is varied and represents a rich collection of pesticides, not only following the most common trends.

\begin{table}[H]
\centering
\begin{tabular}{|c|cc|cc|cc|}
\hline
                 & \multicolumn{2}{c|}{\textbf{Lipinski} \cite{filters_Lipinski}}       & \multicolumn{2}{c|}{\textbf{Hao} \cite{filters_Hao}}            & \multicolumn{2}{c|}{\textbf{Tice} \cite{filters_Tice}}           \\ \hline
\textbf{Dataset} & \multicolumn{1}{c|}{All rules} & 1 violation & \multicolumn{1}{c|}{All rules} & 1 violation & \multicolumn{1}{c|}{All rules} & 1 violation \\ \hline
ApisTox & \multicolumn{1}{c|}{82.6\%}    & 95.1\%      & \multicolumn{1}{c|}{70.9\%}    & 87.7\%      & \multicolumn{1}{c|}{-}         & -           \\ \hline
Herbicides       & \multicolumn{1}{c|}{91.8\%}    & 99.4\%      & \multicolumn{1}{c|}{68.6\%}    & 90.1\%      & \multicolumn{1}{c|}{61.3\%}    & 97.2\%      \\ \hline
Fungicides       & \multicolumn{1}{c|}{88.3\%}    & 94.9\%      & \multicolumn{1}{c|}{79.7\%}    & 90.9\%      & \multicolumn{1}{c|}{-}         & -           \\ \hline
Insecticides     & \multicolumn{1}{c|}{67.0\%}    & 91.6\%      & \multicolumn{1}{c|}{66.1\%}    & 85.0\%      & \multicolumn{1}{c|}{60.8\%}    & 84.1\%      \\ \hline
\end{tabular}
\caption{Percentage of molecules satisfying given chemical rules}
\label{tab:filters}
\end{table}

\subsection*{Molecule structures analysis}

Here, we present analyses concerning internal structure of the molecules, i.e. their Bemis-Murcko scaffolds, functional groups and frequent subgraphs. High quality molecular datasets should, in general, be structurally valid and not be dominated by a few substructures. At the same time, we expect the existence of discriminative structures common only among pesticides toxic or non-toxic for honey bees, which would enable data analysis and interpretability of predictive algorithms.

Firstly, we validate the distribution of molecular scaffolds, which motivated our omission of scaffold split in the \q{Methods} section. Among 1035 molecules, there are 424 Bemis-Murcko scaffolds, and 324 occur in only one molecule, indicating a very diverse dataset. This is also quite high number compared to typical medicinal chemistry datasets. Within the group of toxic molecules, scaffolds composed of 6-membered aromatic rings containing carbons and nitrogens (e.g.\q{c1ccccc1}, \q{c1cncncnc1}, \q{c1ccncc1}, \q{c1ncncn1}) are common, constituting 22\% (67 out of 296) of these molecules.

However, Bemis-Murcko scaffolds are defined only for connected compounds with ring systems, and 186 molecules (almost 20\% of the ApisTox dataset) violate this condition (141 have multiple fragments, 45 do not have any rings). In particular, it means that the ones with disconnected components could share an entire subgraph with those with scaffolds available, and the scaffold split method would not be able to detect that.

\begin{figure}[ht]
\centering
\includegraphics[width=\linewidth]{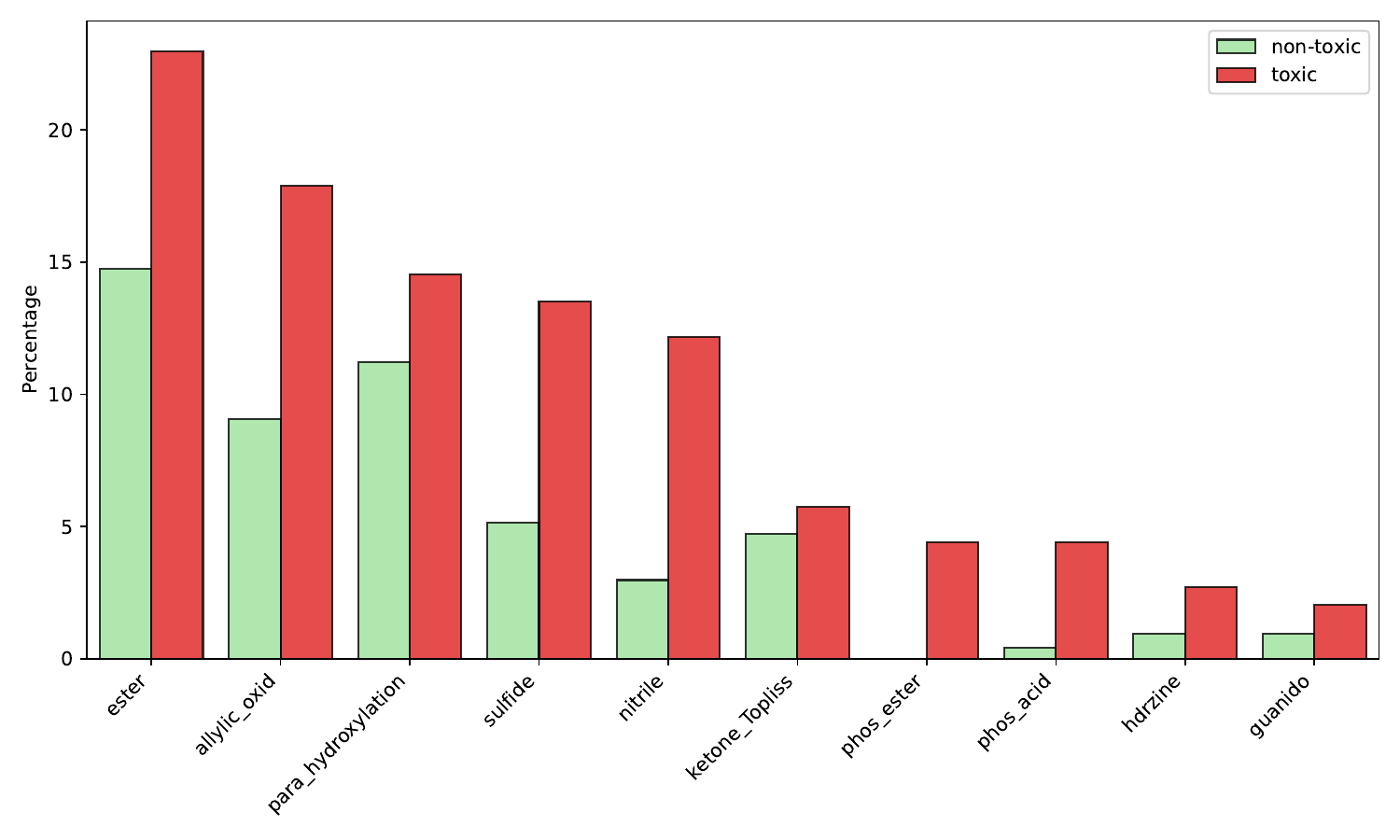}
\caption{Functional groups with the largest frequency difference between non-toxic and toxic molecules. They are sorted by total frequency in the dataset. Groups are named following their corresponding function names in RDKit.}
\label{fig:functional_groups}
\end{figure}

Next, we check the distribution of the functional groups, also called fragments in RDKit. In Figure \ref{fig:functional_groups}, we visualize ten most discriminative functional groups, i.e. those with the largest frequency difference between pesticides toxic and non-toxic to honey bees. We see that there is a high proportion of insecticide-specific functional groups in toxic molecules. Phosphorus is very commonly found in instecticides, most often in a form of organophosphates  \cite{organophosphates_1,organophosphates_2} (\textit{phos\_ester} and \textit{phos\_acid} fragments). Rich presence of \textit{sulfide} fragment is explained by common usage of sulphur in pesticides, in particular fungicides \cite{EPA_sulphur,pesticides_inorganic}. Pyrethrins and pyrethroids constitute a large group of insecticides \cite{pyrethrins_and_pyrethroids}, to which bees are very sensitive \cite{pyrethroids}, containing an ester group and also very often a cyano group (\textit{ester} and \textit{nitrile} fragments). Neonicotinoids are yet another large group of insecticides, very rich in nitrogen, within which a hydrazine group is often detected, as well as a guanidinium group \cite{insecticides1,insecticides2} (\textit{hdrzine} and \textit{guanido} fragments). All those results show that ApisTox data aligns with literature on pesticide toxicity of honey bees.

Lastly, we apply molecular frequent subgraph mining, in order to identify discriminative substructures, i.e. both frequent overall and at the same time much more frequent among toxic molecules than non-toxic ones (or vice versa). This is a more data-driven solution compared to functional groups, since we derive subgraphs from the data itself. It can also detect smaller subgraphs compared to typical functional groups. Again, we expect the data to reflect the usage of common chemical elements and their influence on pesticide toxicity to honey bees.

\begin{figure}[ht]
\centering
\includegraphics[width=\linewidth]{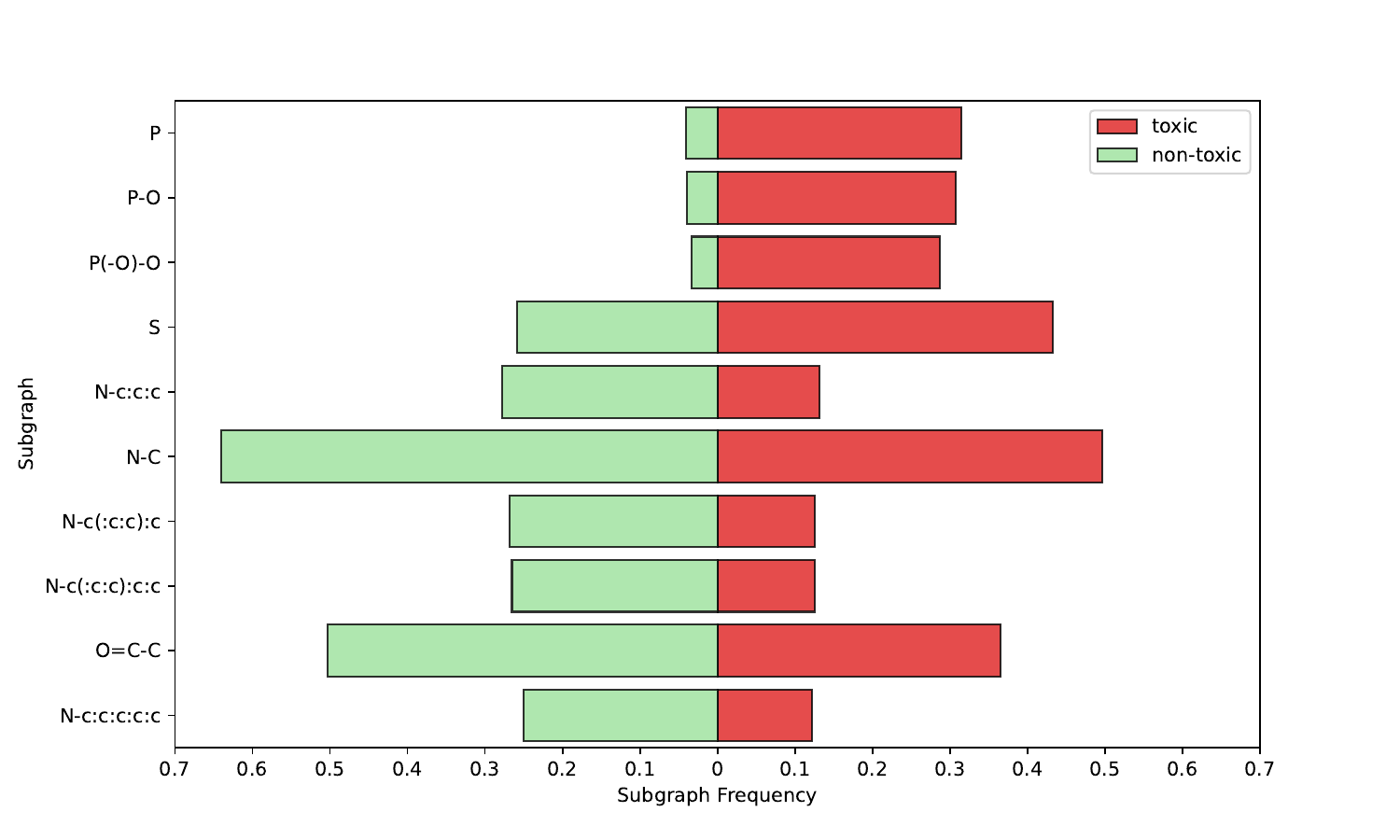}
\caption{Frequency of occurrence of subgraphs identified by MoSS within toxic and non-toxic molecules. We plot groups with the highest frequency difference between classes.}
\label{fig:moss_subgraphs}
\end{figure}

We use the MoSS tool \cite{MoSS} with default settings to mine the most common subgraphs. We then identify their frequency in toxic and non-toxic classes, and select ten with the largest frequency difference between classes. Results are presented in Figure \ref{fig:moss_subgraphs}. The results again align with literature, with phosphate-containing subgraphs much more common among toxic pesticides, and sulfur often found in both types of compounds.

\section*{Usage Notes}

The important application of ApisTox dataset is understanding and predicting agrochemical compounds' toxicity for honey bees, using data mining and machine learning (ML) methods. Those tasks can be divided into unsupervised and supervised applications, depending on whether they explicitly model the dependence between molecule features and selected targets or not.

Analytical tasks include detecting functional groups, scaffolds, binding sites, and other molecular fragments that influence the toxicity of chemicals for honey bees. Those analyses can be carried out either on the entirety of the data, or on its subsets, e.g. for herbicides, fungicides, or insecticides only. Researching those factors can lead to better understanding of chemical space of pesticides non-toxic for honey bees, and therefore to design of safer pesticides in the future. Techniques like graph clustering, dimensionality reduction, and frequent subgraph mining can be utilized \cite{mol_clustering,mol_frequent_subgraph_mining}. We stress that those methods, even when used as unsupervised learning techniques, often have hyperparameters and settings to tune, and they should also be validated on external data not used during the initial analysis. Therefore, the provided train-test splits still have to be used, even if toxicity labels are not utilized. In this regard, time split would be especially useful, as a direct approximation of the process of developing new agrochemicals.

Predictive algorithms will most likely model the relation between structure of molecules (\textit{SMILES} column) and either \textit{label} or \textit{ppdb\_level} column as the target. They can be useful tools for pesticide design, acting as ML-based filters for new agrochemical candidates. Substances with high predicted probability of honey bee toxicity can be ruled out early, lowering costs and accelerating the process. Such models should utilize only the training data for parameter estimation. For hyperparameter tuning, cross-validation with stratified sampling is recommended, due to class imbalance and relatively small dataset size. Test data must not be used at any point until the final validation of generalization performance. We recommend using MaxMin split for this purpose, since it covers the chemical space of the dataset more uniformly than other splits, requiring good generalization across the entire domain of the dataset \cite{split_maxmin,split_maxmin_2}.

For measuring the performance of predictive models, at least two or three metrics should be used, taking into consideration class imbalance and reporting different aspects of model performance. In particular, we recommend Area Under Receiver Operating Characteristic curve (AUROC), since it works well with imbalanced data and also takes into consideration probabilistic outputs of classifiers, as well as Matthews Correlation Coefficient (MCC), which can sometimes detect model failures despite high AUROC value \cite{MCC_recommendation_1,MCC_recommendation_2}. Among other popular metrics, F1-score, precision, and recall can be used.

ApisTox can also be used as a benchmark dataset for molecular graph classification. Since datasets and benchmarks in this area come almost exclusively from medicinal chemistry, performance of many models, like molecular fingerprints, graph kernels, and graph neural networks (GNNs), has been evaluated exclusively on tasks directly related to pharmacological drug design. Our dataset allows validation of the generalization performance of such models on new domains of agrochemistry and ecotoxicology. In this context, usage of unified and predefined split and metrics is of paramount importance, to allow comparison of different models. We recommend usage of MaxMin split, as well as reporting both AUROC and MCC.

When using \textit{ppdb\_level} as target variable, we note that this is not a three class classification problem, but rather ordinal regression, also known as ordinal classification \cite{ordinal_regression}. Toxicity levels are ordered integers, and classes 0 and 2 (non-toxic and highly toxic) are more distant than 1 and 2 (moderately toxic and highly toxic). Therefore, appropriate models should be used, e.g. ordinal logit model instead of logistic regression. In this case, additionally reporting regression metrics like MAE (Mean Absolute Error) and RMSE (Root Mean Squared Error) is recommended, with additional corrections for class imbalance \cite{ordinal_regression_evaluation}.

\section*{Code availability}

Code is available on GitHub at \url{https://github.com/j-adamczyk/ApisTox_dataset}. Code uses Python 3.10. To ensure full reproducibility, we pinned all external library dependencies (including transitive dependencies) using Poetry dependency manager \cite{Poetry}. We also include \textit{poetry.lock} file with all dependency versions, as well as \textit{requirements.txt} file exported from it.

The entire dataset can be recreated from scratch using \textit{create\_dataset.py} script. By default, it uses PPDB and BPDB files from \textit{raw\_data} directory, downloaded at 22nd February 2024, to ensure reproducible results.

\bibliography{bibliography}

\section*{Acknowledgements}

Research project supported by program \q{Excellence initiative – research university} for the AGH University of Krakow.

\section*{Author contributions statement}

J.A. and J.P.: Conceptualization, Investigation, Methodology, Data Curation, Software, Validation, Visualization, Writing - Original Draft, Writing - Review \& Editing. P.S.: Conceptualization, Supervision, Methodology, Writing - Original Draft, Writing - Review \& Editing. All authors reviewed and approved the manuscript.

\section*{Competing interests}

The authors declare no competing interests.

\clearpage

\section*{ApisTox - supplementary information}

\begin{table}[ht]
\begin{tabular}{|c|cc|cc|cc|cc|cc|}
\hline
\textbf{}           & \multicolumn{2}{c|}{\textbf{Herbicides}} & \multicolumn{2}{c|}{\textbf{Fungicides}} & \multicolumn{2}{c|}{\textbf{Insecticides}} & \multicolumn{2}{c|}{\textbf{Other agrochemicals}} & \multicolumn{2}{c|}{\textbf{Unknown}} \\ \hline
\textbf{Split type} & \multicolumn{1}{c|}{Train}     & Test    & \multicolumn{1}{c|}{Train}     & Test    & \multicolumn{1}{c|}{Train}      & Test     & \multicolumn{1}{c|}{Train}         & Test         & \multicolumn{1}{c|}{Train}   & Test   \\ \hline
Stratified random   & \multicolumn{1}{c|}{35\%}      & 32\%    & \multicolumn{1}{c|}{18\%}      & 22\%    & \multicolumn{1}{c|}{22\%}       & 22\%     & \multicolumn{1}{c|}{21\%}          & 21\%         & \multicolumn{1}{c|}{6\%}     & 4\%    \\ \hline
Time                & \multicolumn{1}{c|}{32\%}      & 41\%    & \multicolumn{1}{c|}{19\%}      & 21\%    & \multicolumn{1}{c|}{24\%}       & 15\%     & \multicolumn{1}{c|}{22\%}          & 16\%         & \multicolumn{1}{c|}{5\%}     & 7\%    \\ \hline
MaxMin              & \multicolumn{1}{c|}{37\%}      & 21\%    & \multicolumn{1}{c|}{18\%}      & 21\%    & \multicolumn{1}{c|}{22\%}       & 20\%     & \multicolumn{1}{c|}{18\%}          & 33\%         & \multicolumn{1}{c|}{5\%}     & 6\%    \\ \hline
\end{tabular}
\caption{Pesticide type distributions for different splits.}
\end{table}

Note that pesticides can have more than one type, so it is possible for a given split type for train or test to sum up to more than 100\%.

\end{document}